\documentclass[aps,prd,nofootinbib,twocolumn,superscriptaddress,preprintnumbers,balancelastpage,longbibliography]{revtex4-1}

\usepackage[utf8]{inputenc}
\usepackage{amsmath}
\usepackage{amsfonts} 
\usepackage{amssymb}
\usepackage{graphicx}
\usepackage[dvipsnames]{xcolor}
\usepackage{hyperref}
\usepackage{tensor}
\usepackage{siunitx}
\usepackage{physics}
\usepackage{appendix}
\usepackage{cancel}

\hypersetup{colorlinks=true,
			urlcolor=purple,
			citecolor=blue,
			linkcolor=magenta,}

\newcommand{\beq}{\begin{equation}}
\newcommand{\eeq}{\end{equation}}
\newcommand{\be}{\begin{eqnarray}}
\newcommand{\ee}{\end{eqnarray}}
\newcommand{\bea}{\begin{eqnarray}}
\newcommand{\eea}{\end{eqnarray}}
\newcommand{\bi}{\begin{itemize}}
\newcommand{\ei}{\end{itemize}}
\newcommand{\ben}{\begin{enumerate}}
\newcommand{\een}{\end{enumerate}}
\def\bes{\begin{equation*}}
\def\ees{\end{equation*}}
\def\bead{\begin{aligned}}
\def\eead{\end{aligned}}

\renewcommand{\]}{\right]}
\renewcommand{\(}{\left(}
\renewcommand{\)}{\right)}
\def\bmat{\left(\begin{matrix}}
\def\emat{\end{matrix}\right)}

\def\cL{{\cal L}}
\def\cO{{\cal O}}

\usepackage{mfirstuc} 
\newcommand{\addReviewer}[2]{
  \expandafter\newcommand\csname #1\endcsname[1]{{\textbf{ \color{#2} \capitalisewords{#1}:\,##1}}}
  \expandafter\newcommand\csname #1cor\endcsname[2]{{\color{#2} \capitalisewords{#1}:\,\st{##1}{\textbf{##2}}}}
  \expandafter\newcommand\csname #1color\endcsname{#2}
  \expandafter\newcommand\csname #1todo\endcsname[1]{{\todo[inline,color=white!70!#2, caption={}]{\textbf{\capitalisewords{#1}}: ##1}}}
}

\usepackage{soul,color}
\definecolor{chromeyellow}{rgb}{1.0, 0.65, 0.0}

\begin{document}

\title{A Colorful Mirror Solution to the Strong CP Problem}

\author{Quentin Bonnefoy}
\email{q.bonnefoy@berkeley.edu}
\affiliation{Berkeley Center for Theoretical Physics, Department of Physics, University of California, Berkeley, CA 94720, USA}
\affiliation{Theoretical Physics Group, Lawrence Berkeley National Laboratory, Berkeley, CA 94720, USA}

\author{Lawrence Hall}
\email{ljh@berkeley.edu}
\affiliation{Berkeley Center for Theoretical Physics, Department of Physics, University of California, Berkeley, CA 94720, USA}
\affiliation{Theoretical Physics Group, Lawrence Berkeley National Laboratory, Berkeley, CA 94720, USA}

\author{Claudio Andrea Manzari}
\email{camanzari@lbl.gov}
\affiliation{Berkeley Center for Theoretical Physics, Department of Physics, University of California, Berkeley, CA 94720, USA}
\affiliation{Theoretical Physics Group, Lawrence Berkeley National Laboratory, Berkeley, CA 94720, USA}

\author{Christiane Scherb}
\email{cscherb@lbl.gov}
\affiliation{Berkeley Center for Theoretical Physics, Department of Physics, University of California, Berkeley, CA 94720, USA}
\affiliation{Theoretical Physics Group, Lawrence Berkeley National Laboratory, Berkeley, CA 94720, USA}

\begin{abstract}
We propose theories of a complete mirror world with parity (P) solving the strong CP problem.
P exchanges the entire Standard Model (SM) with its mirror copy. We derive bounds on the two new mass scales that arise: $v'$ where parity and mirror electroweak symmetry are spontaneously broken, and $v_3$ where the color groups break to the diagonal strong interactions. The strong CP problem is solved even if $v_3 \ll v^{\prime}$, when heavy coloured states at the scale $v_3$ may be accessible at LHC and future colliders.
Furthermore, we argue that the breaking of P introduces negligible contributions to $\bar \theta_\text{QCD}$, starting at three-loop order. The symmetry breaking at $v_3$ can be made dynamical, without introducing an additional hierarchy problem.
\end{abstract}

\maketitle


\section{Introduction}

The QCD Lagrangian contains a CP-odd term,
\beq
\bar\theta_\text{QCD} \frac{g_s^2}{32\pi^2}G_a^{\mu\nu}\tilde{G}_{a,\mu\nu}\,,
\label{eq:thetaterm1}
\eeq
where $G_a^{\mu\nu}$ is the gluon field strength tensor, $g_s$ the strong coupling constant, $\tilde{G}_{a,\mu\nu} \equiv \frac{1}{2}\epsilon_{\mu\nu\alpha\beta}G_a^{\alpha\beta}$, and $\bar\theta_\text{QCD}$ is an angle which quantifies the breaking of CP in strong interactions. It is a free parameter of QCD, however experimental constraints on the electric dipole moment of the neutron imply $\bar\theta_\text{QCD} \lesssim 10^{-10}$~\cite{Pendlebury:2015lrz}. The lack of understanding of the smallness of $\bar\theta_\text{QCD}$ has been dubbed the \textit{strong CP problem}. The puzzle is made even sharper by the presence of weak interactions, which are such that $\bar\theta_\text{QCD}$ receives a contribution through the chiral transformation needed to diagonalize the quark mass matrix $M$,
\beq
\bar{\theta}_\text{QCD}=\theta_\text{QCD} +\arg\det(M)
\,.
\label{eq:thetaterm2}
\eeq
where $\theta_\text{QCD}$ is the bare theta angle. The two contributions to $\bar{\theta}_\text{QCD}$ arise from very different physics and have no reason to cancel in the Standard Model.\\
Three approaches to this problem have received considerable attention
in the literature: a massless quark~\cite{Kaplan:1986ru,Banks:1994yg,PhysRevLett.114.141801,Agrawal:2017evu}, spontaneously broken P or CP symmetries~\cite{Babu:1988mw,Babu:1989rb,Barr:1991qx,Nelson:1983zb,Barr:1984qx,Lavoura:1997pq,Vecchi:2014hpa,Dine:2015jga,Hall:2018let,Dunsky:2019api,Craig:2020bnv}, or a spontaneously broken global chiral symmetry à la Peccei-Quinn \cite{Peccei:1977hh,Peccei:1977ur,Weinberg:1977ma,Wilczek:1977pj}. 
\\
Although it was recognized in the 1970s that parity might solve the strong CP problem~\cite{Beg:1978mt, Mohapatra:1978fy}, early attempts to construct such theories, based on the Left-Right extension of the electroweak group to $SU(2)_L \times SU(2)_R \times U(1)_{B-L}$, were unsuccessful, until Babu and Mohapatra discovered a solution in a simple model with a separate Higgs doublet for each $SU(2)$ group~\cite{Babu:1988mw}. In this model, P forces $\theta_\text{QCD}$ to zero and the fermion Yukawa matrices to be Hermitian, hence $\bar{\theta}_\text{QCD}=0$ at tree-level. Shortly after, the same authors UV-completed their construction through a see-saw mechanism involving heavy vector-like fermions~\cite{Babu:1989rb}. A realistic vacuum can occur at tree-level, via a soft breaking of P, or can arise radiatively~\cite{Hall:2018let}. In both cases, the resulting radiative corrections to $\bar{\theta}_\text{QCD}$ occur at 2-loop order and can be small enough to solve the strong CP problem, while offering the prospect of an observable neutron electric dipole moment~\cite{deVries:2021pzl,Hall:2018let, Hisano:2023izx}.
An alternative model was constructed in Ref.~\cite{Barr:1991qx} by doubling the SM electroweak group to
$(SU(2)_L \times U(1)_Y) \times (SU(2)’ \times U(1)’)$. A single SM-like Higgs doublet and three generations of SM-like fermions were introduced for each of the SM and mirror electroweak sectors. 
Both sectors share a common strong interaction. P forces $\theta_\text{QCD}$ to zero, while the Yukawa matrices for SM fermions are arbitrary but hermitian conjugates of those of mirror fermions, so that the quark contributions to $\bar{\theta}_\text{QCD}$ are cancelled by the mirror contributions. A hierarchy of Higgs vevs, $v’ \gg v$, can again be obtained either by soft P breaking or by radiative contributions to the Higgs potential~\cite{Dunsky:2019api}; in both cases the contributions to $\bar{\theta}_\text{QCD}$ arise at three loops, as with radiative contributions in the SM~\cite{Ellis:1978hq}, and are small.  In a final comment of Ref.~\cite{Barr:1991qx}, it was suggested that this theory could be unified, into an $SU(5) \times SU(5)^{\prime}$ or $SO(10) \times SO(10)^{\prime}$ theory. The resulting $SU(3) \times SU(3)^{\prime}$ group of strong interactions would be reduced to QCD by breaking to the diagonal $SU(3)_{\rm QCD}$ combination, with P then fixing the $\bar \theta$ parameter of QCD to zero. Such unified theories have not been constructed, and we are not aware of any discussions of this mechanism in the literature.\\
In this paper, we propose the simplest theory of a complete mirror world with P solving the strong CP problem, and discuss its phenomenology. Parity exchanges the entire SM for its mirror copy and there are only two relevant free parameters beyond those of the SM. One is the mass scale $v'$, where parity and mirror electroweak symmetry are spontaneously broken, and the other is the mass scale $v_3$, where $SU(3) \times SU(3)^{\prime}$ breaks to $SU(3)_{\rm QCD}$. Importantly, the strong CP problem is solved even if $v_3 \ll v^{\prime}$.

\section{Solution to the Strong CP Problem}
\label{sec:Solution}

We mirror the full SM gauge group and therefore start with $SU(3)\times SU(2)_L\times U(1)_Y \times SU(3)^{\prime}\times SU(2)^{\prime}\times U(1)^{\prime}$. In Left-Right theories, one leaves the fermion content of the SM unchanged, and the right-handed (RH) fermions are grouped into prime electroweak doublets. Here, however, we cannot simply assign prime color to such a doublet, as that would generate $SU(3)^{(\prime)}{}^3$ gauge anomalies. Therefore, we are led to a complete mirror theory, where one also doubles the fermionic spectrum. We denote the mirror partners of the SM fields with a prime. In particular, we have two Higgs bosons, each charged under only one of the two worlds and responsible for the breaking of its electroweak sector. Our mirror world Lagrangian thus reads
\beq
{\cal L}={\cal L}_\text{SM} + {\cal L}_{\text{SM}’}+\tilde\lambda\abs{H}^2\abs{H’}^2 \ ,
\eeq
where ${\cal L}_{\text{SM}'}$ has the same form as the SM Lagrangian, but all fields and couplings are primed.
For simplicity, the kinetic mixing\footnote{Note that the kinetic mixing can be forbidden at tree-level by embedding the $U(1)$ gauge groups into a larger non-abelian group. Moreover, the loop contributions below the breaking scale of that new group are naturally suppressed in our model. Finally, collider bounds on the model given an $\cO(1)$ kinetic mixing are weaker than the limits considered in the following and we do not consider cosmological ones, as the latter depend on assumptions on the cosmological history which we remain agnostic about in the present paper and leave for future work.} and dimension-5 neutrino mass operators are not shown, as we do not need them for our analysis of the strong CP problem.
\\
The gauge and field content of the model is now such that one can pair the fields via spacetime parity. More precisely, we compose the usual action of P 
with a $\mathbb{Z}_2$ symmetry which exchanges the SM and mirror fields. At the level of the gauge bosons, one has
\beq
\label{eq:CPtransformGaugeBosons}
A^a_\mu(t,\vec r) \xrightarrow[]{\mathbb{Z}_2} A'^a_\mu(t,\vec r)  \xrightarrow[]{P} A'^a{}^\mu(t,-\vec r) \ .
\eeq
(Due to the independent C invariance of each Yang-Mills theory, even with $\theta$ terms, it is actually equivalent to impose either P or CP. See the Supplemental Material at the end of that document. For simplicity in the following we impose P.) Parity interchanges left-handed (LH) and RH fermions, and hence sends
\begin{equation}
Q(t,\vec r) \xrightarrow[]{P\circ \mathbb{Z}_2} \gamma^0 Q^{\prime\, c}(t,-\vec r)\,,
\label{eq:PPlusZ2fermions}
\end{equation} 
with $Q^{\prime}$ in the $\overline{\mathbf{3}}$ of $SU(3)^{\prime}$, and similarly for the other fermions. It also exchanges $H\leftrightarrow \tilde{H'}$. The $\theta$ angles are odd under P, hence $\theta = -\theta^{\prime}$ and $\cL$ only contains
\beq
\begin{split}
\frac{\theta}{16\pi^2}\[g_3^2\Tr\(G\tilde G\)-g_3'^2\Tr\(G'\tilde G'\)\] \, .
\label{eq:PInvarianceTheta}
\end{split}
\eeq
Finally, parity imposes that the SM and mirror gauge couplings are equal, in particular $g_3=g_3^{\prime}$, but we keep the two couplings explicit in \eqref{eq:PInvarianceTheta} since they will run differently below the scale of P breaking. From \eqref{eq:PInvarianceTheta}, it appears clearly that breaking $SU(3)\times SU(3)'$ to its diagonal subgroup (then identified with $SU(3)_{\rm QCD}$) provides a perfect cancellation of $\bar \theta_{\rm QCD}$, as long as no new phases are introduced by the sector that generates this breaking. This discussion accounts also for the electroweak contributions: P sends $\bar Q u\tilde H$ to $\(\bar Q' u'\tilde{H'}\){}^\dagger$, hence the Yukawa matrices in the SM and mirror sectors are hermitian conjugates of one another. 
Thus, the $\bar\theta^{(\prime)}$ angles also appear as in Eq.~\eqref{eq:PInvarianceTheta}. We therefore extend the Lagrangian of our model
in order to accommodate the color breaking:
\begin{equation}
    \mathcal{L} \to \mathcal{L} + \mathcal{L}_\text{\rm breaking} \ .
\end{equation}
We discuss $\mathcal{L}_\text{\rm breaking}$ in more detail below.

\section{$SU(3)\times SU(3)^{\prime}$ breaking}
\label{sec:ScalarSector}

The mechanism presented in Sec~\ref{sec:Solution} does not depend on a specific symmetry breaking sector, $\mathcal{L}_\text{\rm breaking}$, as long as it provides the breaking pattern $SU(3)\times SU(3)^{\prime} \to SU(3)_{\rm QCD}$. We present here a simple realization of this symmetry breaking by the vev of a bi-fundamental scalar field $\Sigma_{ii'}$. We nevertheless stress that this realization is by no means unique and different scenarios have different particle spectra and phenomenological signatures. For instance, models that dynamically break the color groups to their diagonal subgroup are discussed in the Supplemental Material below.\\
Modulo conjugation, we can consider two cases for the charges of $\Sigma$: $(\mathbf{3},\mathbf{3})$ or $(\mathbf{3},\overline{\mathbf{3}})$. When $\Sigma$ acquires a vev proportional to the identity matrix, $\langle\Sigma\rangle=\frac{v_3}{\sqrt 6}\mathbf{1}$, the vacuum preserves the diagonal $SU(3)_\text{QCD}$.
Projecting onto the massless gluons, one finds $g_3^2G\tilde G=g_3'^2G'\tilde G'$, confirming the cancellation of $\theta_\text{QCD}$ from Eq.~\eqref{eq:PInvarianceTheta} (see the Supplemental Material for details). Anticipating the discussion in Sec.~\ref{sec:EnergyScales}, we stress that this cancellation holds even when $g_3\neq g_3'$, as long as $\theta= -\theta'$. Such a vev is easy to achieve via the most general potential of $\Sigma$, which reads
\beq
\begin{split}
V(\Sigma) =& -m^2\Tr(\Sigma\Sigma^\dagger)+c\Tr^2(\Sigma\Sigma^\dagger)\\
&+\tilde c\Tr(\Sigma\Sigma^\dagger)^2+\(\tilde m\det(\Sigma)+h.c.\)
\ .
\end{split}
\label{eq:SigmaPotential}
\eeq
In addition to this potential, $\cL_{\rm breaking}$ contains mixing terms between $\Sigma$ and the Higgs fields. Since $v_3\gg v$, the couplings to $H$ are irrelevant for our discussion, while the vev of $H'$ simply shifts the couplings shown in Eq.~\eqref{eq:SigmaPotential}.
The vacua of this model have been thoroughly studied in Ref.~\cite{Bai:2017zhj}: there are parameter ranges where the unbroken gauge symmetry in the global minimum is $U(1)^2$, $SU(2)^2\times U(1)$ or $SU(3)$. Only the latter is of interest for us, which is for instance the only (global and local) minimum when $m^2\geq 0$ and $c,\tilde c\geq 0$ (see Ref.~\cite{Bai:2017zhj} for the complete set of conditions).
Finally, we stress that with our choice of charges, $\Sigma$ does not couple to fermions at the renormalizable level and its vev does not reintroduce CP phases in their mass matrices.\\
We discuss the need for parity breaking in the next section, however here we note that if $v_3$ is larger than the parity breaking scale, the low energy description of our model coincides with models where only electroweak forces are mirrored \cite{Babu:1988mw,Babu:1989rb,Barr:1991qx,Hall:2018let,Dunsky:2019api,Craig:2020bnv}. 
On the other hand, experimental bounds are much stricter on the scale of parity breaking than they are on $v_3$, hence the scenario where $v_3$ is at the lowest possible scale is the most phenomenologically interesting and novel. Further details on the associated spectrum of physical scalars in the IR are given in the Supplemental Material.

\section{Parity Breaking and\\Energy Scales}
\label{sec:EnergyScales}

\begin{figure}[tb]
	\centering
	\includegraphics[width=1.\columnwidth]{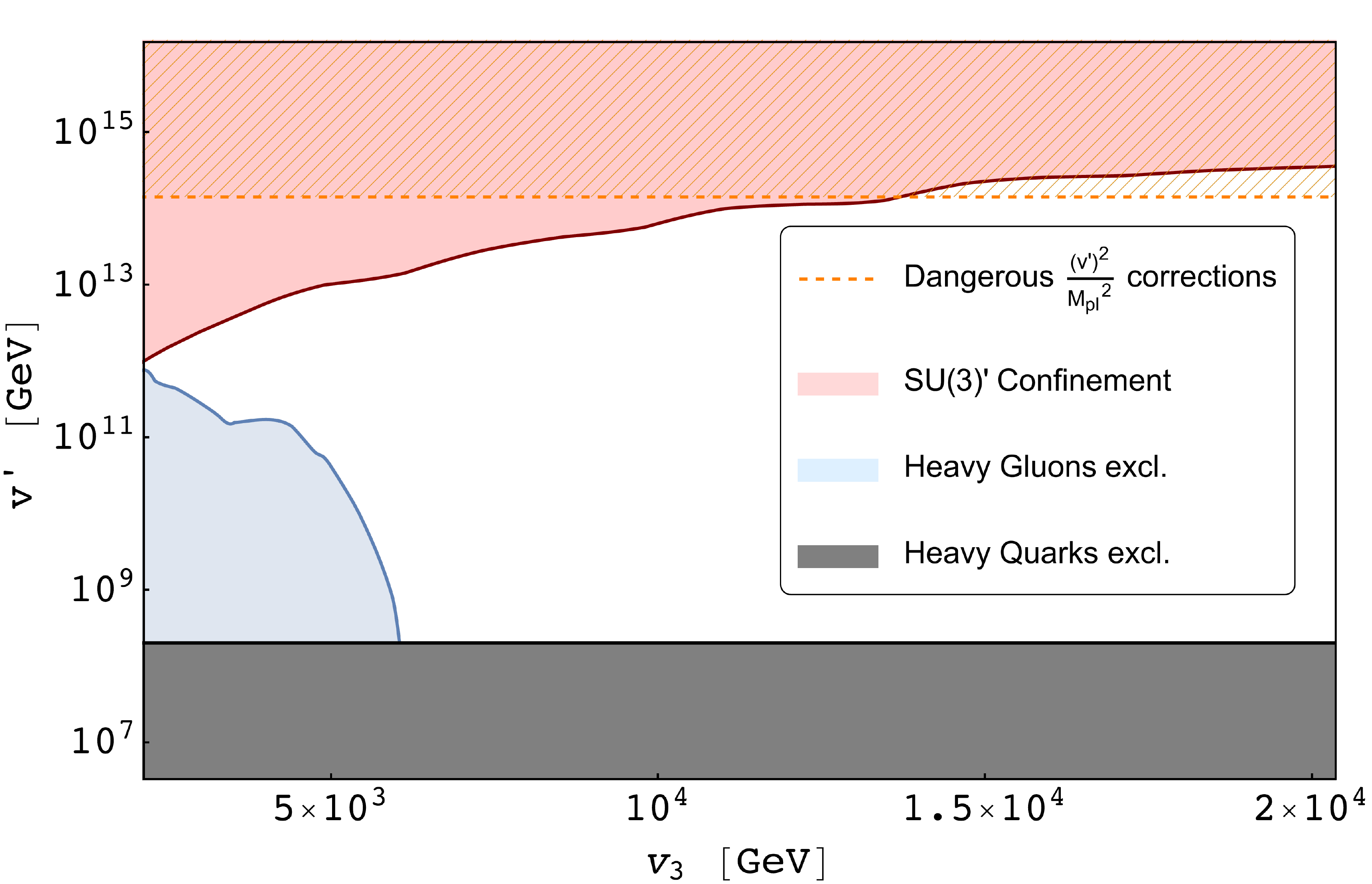}
	\caption{Allowed parameter space in the $v_3$-$v^{\prime}$ plane. The region excluded by collider searches for heavy gluons (blue) and heavy quarks (gray), by the requirement that $SU(3)^{\prime}$ does not confine before $v_3$ (solid red line) and that higher dimensional operators do not reintroduce a sizable $\theta_{\rm QCD}$ (dashed orange line) are shown.}
	\label{fig:v3vpPlot}
\end{figure}

We showed in the previous sections how to obtain a perfect cancellation of $\bar\theta_\text{QCD}$ when P connects the SM to its mirror copy. However, if P is unbroken, this possibility is ruled out by experiment.
Collider and cosmological probes require the mirror sector to decouple at low energies, and therefore P must be broken at some high energy scale. This is most easily achieved by making the vev of the mirror Higgs much larger than that of its SM companion: $v^{\prime} \gg v$. Such a hierarchical vacuum can be obtained at tree-level, via explicit soft breaking of parity \cite{Babu:1988mw,Babu:1989rb,Craig:2020bnv}, or through loop-induced corrections, as in Higgs Parity~\cite{Dunsky:2019api}. These mechanisms for spontaneous breaking of parity in the electroweak sector can occur even when the Higgs doublets have quartic couplings to the colored $\Sigma$ field, regardless of whether $v_3$ is larger or smaller than $v'$. In any case, since the parity breaking is spontaneous or soft, the mirror Yukawa matrices remain the hermitian conjugates of those of the SM at the scale $v'$. The present solution to the strong CP problem is therefore completely defined by two energy scales: $v^{\prime}$ and $v_3$. The additional parameters associated with the specific $SU(3)\times SU(3)^{\prime}$ breaking mechanism are not relevant for the strong CP problem, as long as they provide the right breaking pattern (but they are relevant for studying the phenomenology of any precise model).\\
Independently of the breaking mechanism, there is the requirement that the model solves the strong CP problem, despite parity being broken. Contributions to $\bar \theta_\text{QCD}$ beyond those of Section~\ref{sec:Solution}, which cancel, can be classical or quantum. Quantum contributions to $\bar\theta_\text{QCD}$ are discussed in the next section. Classical ones yield an upper bound on $v^{\prime}$ due to expected new physics at most at the Planck scale. More precisely, there are dangerous dimension-6 operators of the form
\beq
\frac{g_3^2\Tr G\tilde G}{16\pi^2M_P^2}\(\lambda \abs{H}^2 +\lambda' \abs{H'}^2\)-\(g_3,G,H\leftrightarrow g_3',G',H'\) \ ,
\eeq
where the pattern of couplings is chosen so as to respect parity~\cite{PlanckOp}. There can also be corrections to the Yukawa couplings of the form
\beq
\bar Q\frac{Y_{u,1}\abs{H}^2 +Y_{u,2}\abs{H'}^2}{M_P^2}u \tilde H
+\(\begin{matrix}Y_{u,i},Q,u,H\\\leftrightarrow Y_{u,i}^\dagger,Q',u',H'\end{matrix}\) \ ,
\eeq
and similarly in the down sector. When the two Higgses receive different vevs, such operators reintroduce $\bar\theta_\text{QCD}\neq 0$. For order one Wilson coefficients, the presence of the first kind of operators imposes $v^{\prime}\lesssim 10^{14}\, {\rm GeV}$, as shown in Fig.~\ref{fig:v3vpPlot}. The contribution of the second kind is enhanced by the inverse of the small quark Yukawas, strengthening the bound by roughly $2.5$ orders of magnitude, unless the flavor structure of the matrices $Y_{u,i}$ is similar to that of $Y_u$ in a full model of flavor. In that case, the bound is unchanged.\\
A second set of model-independent constraints comes from collider bounds on the mirror quarks, which become charged under $SU(3)_\text{QCD}$ below $v_3$. The lightest mirror quarks, in particular the mirror up-quark, are stable and, once pair-produced in p-p collisions, they form fractionally-charged colorless bound states with SM quarks or gluons produced by their own color field~\cite{Fairbairn:2006gg}. Therefore, the best constraints come from LHC searches for heavy stable electrically charged particles~\cite{CMS:2021pmn,ATLAS:2022pib}. We recasted the ATLAS search for stable gluinos and charginos of Ref.~\cite{ATLAS:2022pib}, finding a lower bound of $m_{u^{\prime}}\gtrsim 1.3\; \rm TeV$. The Yukawas of the two worlds being related by parity, this translates into $v^{\prime}\approx \sqrt 2 m_{u^{\prime}}/Y_u \gtrsim 1.5\times 10^8\; \rm GeV$, as shown in Fig.~\ref{fig:v3vpPlot}.\\
Further constraints come from the different runnings of the gauge couplings and the presence of additional fields. While the former are qualitatively independent of the $SU(3)\times SU(3)'$ breaking mechanism, the latter are strongly model-dependent. The running of $g_3$ differs from that of $g_3^{\prime}$ below $v^{\prime}$, since the quarks are much heavier in the mirror sector~\cite{g3pHDO}. As we anticipated in Section~\ref{sec:ScalarSector}, this does not affect the cancellation of $\bar\theta_\text{QCD}$.  Nevertheless, we ought to require that $SU(3)^{\prime}$ does not confine before $v_3$, as it would modify the potential of the order parameter breaking $SU(3)\times SU(3)^{\prime}$. The confinement of one of the two gauge groups does not ensure anymore that the two groups are broken to the diagonal subgroup and that the strong CP problem is solved. Since the gauge coupling running depends on the precise particle content, we consider for concreteness the case of the bi-fundamental scalar discussed in Sec.~\ref{sec:ScalarSector}. We then obtain another upper bound for $v^{\prime}$ as a function of $v_3$, as shown in Fig.~\ref{fig:v3vpPlot}. 
Note that the gauge couplings must match to the QCD coupling constant at $v_3$ and be equal at $v'$, hence their values and runnings are known for given $v'$ and $v_3$.\\
The spectrum of the theory contains a massive color-octet vector coupled to (mirror) quarks and various scalar states, some of which are charged under $SU(3)_\text{QCD}$ (see the Supplemental Material for further details). Bounds from collider searches of the former have been extensively discussed in Refs.~\cite{Dobrescu:2007yp,Bai:2018jsr} and are shown in Fig.~\ref{fig:v3vpPlot}.  
Accounting for the  requirement that $SU(3)^{\prime}$ does not confine, we find $v_3 \gtrsim 0.85\, \text{\rm TeV}$. For these values, the collider bounds on coloured scalars can always be avoided by an appropriate choice of the parameters in the $\Sigma$ potential~\cite{Bai:2018jsr}, so we do not discuss them further. In other realizations of the color breaking mechanism, there may be scalars or fermions significantly lighter than the heavy gluons, resulting in the blue region extending towards the right side of the plot.

\section{Quantum corrections to $\bar \theta$}
\label{sec:thetaRunning}

The fact that all classical sources of strong CP violation (CPV) cancel to sufficient accuracy does not yet ensure that the model solves the strong CP problem: one also needs to check that the spontaneous breaking of parity does not reintroduce $\bar \theta_\text{QCD}$ at the quantum level~\cite{Ellis:1978hq,deVries:2018mgf,Valenti:2021rdu,deVries:2021sxz,deVries:2021pzl}.
We find that, in our model, no contributions to $\bar \theta_\text{QCD}$ exist before three-loop
 order.\\
To see this, it is useful to remind that one needs both P and CP violation to generate a non-zero $\bar \theta_\text{QCD}$.
To begin with, the vev of $\Sigma$ does not spontaneously break (C)P and does not introduce any new CP phase. Indeed, for real $\tilde m$ (which can always be achieved upon rephasing $\Sigma$), $\langle\Sigma\rangle$ can be chosen to be diagonal and real \cite{Bai:2017zhj}. Therefore, the gauge, self-interactions and vev of $\Sigma$ respect the various discrete symmetries which act as follows when $\Sigma$ transforms as a $\(\mathbf{3},\mathbf{\bar 3}\)$ (see the Supplemental Material for further details),
\beq
\label{eq:sigmaTransf}
\Sigma\xrightarrow[]{P\circ\mathbb{Z}_2}\Sigma^\dagger\xrightarrow[]{C}\Sigma^T \ .
\eeq
In particular, a real $\tilde m$ is compatible with P. Similarly, the vevs of the two Higgses can both be chosen real via gauge transformations, hence neither the scalar potential nor the scalar vevs break CP. Therefore, the physical sources of CPV are fully contained in the Yukawa matrices and are constrained by the large $U(3)^{6}$ quark flavor symmetry: they reduce to two copies of the Jarlskog invariant of the SM \cite{Jarlskog:1985ht, Jarlskog:1985cw}. At energies larger than $v'$, parity equates them 
 while below $v'$, parity is broken and they run differently. Nevertheless, there are no diagrams contributing to $\bar \theta_\text{QCD}$ before the three-loop order. The argument goes as follows: corrections to $\bar \theta_\text{QCD}$ are associated to loop corrections to the two-point functions of fermions, while Jarlskog-like structures only arise in diagrams with at least four CKM insertions. It is quite simple to see that the simplest diagrams arise at two-loops, with two W boson propagators closing onto a single quark line. However, those involve a single SM copy at a time and it has been shown in Ref.~\cite{Ellis:1978hq} that they vanish~\cite{2loopthetaP}. In our model, it also turns out that there are very few three-loop diagrams beyond those already considered by Ref.~\cite{Ellis:1978hq}. New diagrams would either mix the two SM copies, or involve the new boson $\Sigma$. Above the scale $v_3$, both kind of diagrams require four W bosons, and at least two gluon lines in addition to a line of either $\Sigma$ or a mirror fermion, or a vertex mixing $H,H'$. They are all at least four-loop suppressed. Below $v_3$, a single kind of new three-loop diagrams exists, namely the exact copies of the leading diagrams considered in Ref.~\cite{Ellis:1978hq}, upon replacing massless gluon propagators by massive ones and $g_s\to (g_3/g_3')^{\pm 1}g_s$ for each of the two copies. Since our massive gluons are heavy and $g_3\sim g_3'$, their contributions is at most comparable to that of massless gluons. We therefore conclude that loop contributions to $\bar\theta_\text{QCD}$ in our model are comparable to those in the SM, and totally negligible. Let us stress that, although we explicitly referred to $\Sigma$ in the previous discussion, the conclusion also holds for composite models, discussed in the Supplemental Material.
Finally, there are non-perturbative contributions to $\bar\theta_\text{QCD}$ which are sensitive to the non-zero CP-odd $\theta$ angles above $v_3$. Despite being non-perturbative, those effects can be sizeable due to the fact that the gauge couplings of $SU(3)\times SU(3)'$ can be much larger than that of $SU(3)_\text{QCD}$ at $v_3$ \cite{Affleck:1980mp,Csaki:1998vv,Agrawal:2017evu,Agrawal:2017ksf,Csaki:2019vte}.
The small instantons generate fermionic 't Hooft determinants \cite{tHooft:1976snw} which lead to corrections to the fermion masses. As said above, the absence of free parameters beyond $v_3$ and $v'$ in our model allows us to compute those terms unambiguously. Those involving the SM fermions are sufficiently suppressed by the product of all Yukawa couplings, while those involving the mirror fermions are suppressed by the small gauge couplings when $v'$ is large. They can become sizeable when the mirror up quark mass is close to $v_3$, but such situations also correspond to gauge couplings which remain small at $v_3$, as can be seen from Fig.~\ref{fig:v3vpPlot}. We have checked that the induced shift of $\bar\theta_\text{QCD}$ is compatible with the current bounds, and does not exclude more regions of parameter space in Fig.~\ref{fig:v3vpPlot}.\\

\section{Conclusion}
\label{sec:Conclusions}

We have proposed simple theories of a complete mirror world where parity composed with the mirror exchange symmetry solves the strong CP problem. The new feature of our construction is the presence of a mirror strong interaction, and therefore of two non-vanishing $\theta$ angles. The solution to the strong CP problem and the experimental viability of the model rely on two symmetry breakings: the breaking of the color groups to their diagonal subgroup at the scale $v_3$ makes the effective low-energy $\bar \theta_\text{QCD}$ angle vanish through destructive interference, while the coloured mirror fermions are made heavy through the breaking of parity by a large mirror electroweak scale $v' \gg v$. We focus on the scenario where $v_3\ll v'$;  if $v_3\geq v'$, the effective theory below $v_3$ is the model of Ref.~\cite{Barr:1991qx, Dunsky:2019api}. In addition, saturating the experimental constraints allows $v_3$ to be much below $v'$, leading to the richest phenomenology; new colored states may be accessible at colliders. Due to the symmetry structure of the model, the loop corrections to $\bar \theta_\text{QCD}$ are shown to be under control everywhere in parameter space. We stress the high predictive power of this mirror world, with only two scales characterizing its qualitative features: $v_3$ and $v^{\prime}$. All the other scales in the mirror world are related by parity to those in the SM. It is also worth noting the rich cosmology of our models due to the presence of heavy fermions, scalars and vectors at different energy scales, as well as various phase transitions. These topics are currently under investigation.

\section*{Acknowledgments}
We thank the members of the Berkeley Center for Theoretical Physics for useful discussions. We also thank Simon Knapen, Hitoshi Murayama, Michael Peskin and Pablo Qu\'ilez for discussions on composite models, Simon Knapen and Dean Robinson for discussions on collider bounds, and Pablo Qu\'ilez for suggesting that we check the contributions of small instantons. We are deeply grateful to Simone Pagan Griso for guiding us through the appropriate LHC searches. This work is supported by the Office of High Energy Physics of the U.S. Department of Energy under contract DE-AC02-05CH11231 and by the NSF grant PHY-2210390. CS acknowledges additional support through the Alexander von Humboldt Foundation.


\bibliography{biblio}

\begin{thebibliography}{47}
\bibitem [{g3p()}]{g3pHDO}%
  \BibitemOpen
  \href@noop {} {}\bibinfo {note} {There are also higher-dimensional terms
  which lift the degeneracy between the $SU(3)^{(\prime)}$ gauge couplings
  below $v'$ at the classical level, \bes \frac{\Tr
  G^2}{M_P^2}\(c\abs{H}^2+c'\abs{H'}^2\)+\(G,H,g_3\leftrightarrow G',H',g_3'\)
  \ . \ees Due to the small $v'/M_P$ ratio, such effects are negligible with
  respect to those from RG running.}\BibitemShut {Stop}%
\end{thebibliography}

\end{document}


\title{Supplemental Material to\\
``A Colorful Mirror Solution to the Strong CP Problem"}

\maketitle

\section{Fields and symmetries}

The matter fields of the models have the gauge charges shown in Table~\ref{tab:Spectrum}.
\begin{table}[h!]
    \centering
    \begin{tabular}{c|c|c|c|c|c|c}
        & $SU(3)$ & $SU(2)_L$ & $U(1)_Y$& $SU(3)'$ & $SU(2)'$ & $U(1)'$  \\
        \hline
        $Q$ & $\mathbf{3}$ & $\mathbf{2}$ & 1/6 & $\mathbf{1}$ & $\mathbf{1}$ & 0 \\  
        $u^c$ & $\overline{\mathbf{3}}$ & $\mathbf{1}$ & -2/3 & $\mathbf{1}$ & $\mathbf{1}$ & 0 \\
        $d^c$ & $\overline{\mathbf{3}}$ & $\mathbf{1}$ & 1/3 & $\mathbf{1}$ & $\mathbf{1}$ & 0 \\
        $L$ & $\mathbf{1}$ & $\mathbf{2}$ & -1/2 & $\mathbf{1}$ & $\mathbf{1}$ & 0 \\
        $e^c$ & $\mathbf{1}$ & $\mathbf{1}$ & -1 & $\mathbf{1}$ & $\mathbf{1}$ & 0 \\
        $H$ & $\mathbf{1}$ & $\mathbf{2}$ & 1/2 & $\mathbf{1}$ & $\mathbf{1}$ & 0 \\
        \hline
        $Q'$ & $\mathbf{1}$ & $\mathbf{1}$ & 0 & $\overline{\mathbf{3}}$ & $\mathbf{2}$ & -$1/6$ \\  
        $u'^c$ & $\mathbf{1}$ & $\mathbf{1}$ & 0 & $\mathbf{3}$ & $\mathbf{1}$ & $2/3$ \\
        $d'^c$ & $\mathbf{1}$ & $\mathbf{1}$ & 0 & $\mathbf{3}$ & $\mathbf{1}$ & -$1/3$ \\
        $L'$ & $\mathbf{1}$ & $\mathbf{1}$ & 0 & $\mathbf{1}$ & $\mathbf{2}$ & $1/2$ \\
        $e'^c$ & $\mathbf{1}$ & $\mathbf{1}$ & 0 & $\mathbf{1}$ & $\mathbf{1}$ & $1$ \\
        $H'$ & $\mathbf{1}$ & $\mathbf{1}$ & 0 & $\mathbf{1}$ & $\mathbf{2}$ & -$1/2$ \\
    \end{tabular}
    \caption{Left-handed (LH) fields of the model, related by parity as shown in Eq.~\eqref{eq:parityTransfos}
    .}
    \label{tab:Spectrum}
\end{table}

Including the gauge fields, they are related by parity according to
\beq
\label{eq:parityTransfos}
\(\begin{matrix}
    A^a_\mu(t,\vec r)\\
    H(t,\vec r)\\
    Q(t,\vec r)
\end{matrix}\) \xrightarrow[]{P\circ \mathbb{Z}_2}
\(\begin{matrix}
    A'^a{}^\mu(t,-\vec r)\\
    H'(t,-\vec r)\\
    \gamma^0 Q^{\prime\, c}(t,-\vec r)
\end{matrix}\) \ ,
\eeq
and similarly for the other fermions.

In the main text, we have also introduced a scalar field $\Sigma$, which transforms as a bi-fundamental of the color groups in the $(\mathbf{3},\mathbf{3})$ or $(\mathbf{3},\mathbf{\bar 3})$ representation. (We discuss the cases where the breaking of the color groups is dynamical below.) Calling $U,U^{\prime}$ the unitary group actions in the $\mathbf{3}$ representation, one has
\beq
\Sigma\left\{\begin{matrix} \xrightarrow[]{(\mathbf{3},\mathbf{3})} U\Sigma U'^T \\
\xrightarrow[]{(\mathbf{3},\overline{\mathbf{3}})} U\Sigma U'^\dagger 
\end{matrix}\right.
\eeq
such that the kinetic term $\Tr\(\(D_\mu\Sigma\]^\dagger D_\mu\Sigma\)$ of $\Sigma$ is built from
\beq
\begin{split}
(\mathbf{3},\overline{\mathbf{3}}) \quad & D_\mu \Sigma=\partial_\mu\Sigma-ig_3T^aG^a_\mu\Sigma+ig_3'\Sigma T^aG'^a_\mu\,,\\
(\mathbf{3},\mathbf{3})\quad &D_\mu \Sigma=\partial_\mu\Sigma-ig_3T^aG^a_\mu\Sigma-ig_3'\Sigma T^a{}^TG'^a_\mu\,.
\end{split}
\eeq
This imposes the following choice
\beq
\begin{split}
(\mathbf{3},\overline{\mathbf{3}}) \quad&\Sigma\xrightarrow[]{P\circ\mathbb{Z}_2}\Sigma^\dagger\xrightarrow[]{C}\Sigma^T\,,\\
(\mathbf{3},\mathbf{3})\quad &\Sigma\xrightarrow[]{P\circ\mathbb{Z}_2}\Sigma^T\xrightarrow[]{C}\Sigma^\dagger\,.
\end{split}
\eeq
These transformations restrict a class of higher-dimensional operators which could generate a non-zero $\bar\theta_\text{QCD}$, $\Tr\(\Sigma G' \Sigma^\dagger \tilde G\)$ where $G^{(\prime)}\equiv G^{(\prime)a}T^a$. Such an operator is real, and it is forbidden by parity if $\Sigma$ transforms as a $(\mathbf{3},\overline{\mathbf{3}})$. When $\Sigma$ transforms as a $(\mathbf{3},\mathbf{3})$, one can write $\Tr\(\Sigma G' \Sigma^\dagger \tilde G\)-\(G^{(\prime)}\to G^{(\prime)*}\)$, which cancels when $\Sigma$ is replaced by its vev. This is consistent with the fact that $\langle\Sigma\rangle$ does not break P nor CP.

\section{P or CP}
\label{app:PCP}

Due to the independent C invariance of each Yang-Mills theory (even with $\theta$ terms), the strong CP problem can be solved imposing either P or CP. In fact, with one Weyl fermion, CP is the only well-defined transformation, whereas the difference between P and CP is blurred when one considers two mirror Weyl fermions and insists on exchanging them. In the main text we made the choice of imposing P, while here we discuss CP.\\
The action of (C)P on the gauge fields $A_\mu^a$, when composed with a $\mathbb{Z}_2$ symmetry which exchanges the SM and mirror fields, would yield
\beq
\begin{split}
A^a_\mu(t,\vec r) \xrightarrow[]{\mathbb{Z}_2} A'^a_\mu(t,\vec r)  &\xrightarrow[]{P} A'^a{}^\mu(t,-\vec r)\\
&\xrightarrow[]{C} -s^aA'^a{}^\mu(t,-\vec r) \ ,
\end{split}
\eeq
where $T^a{}^*=s^aT^a$ for $s^a=\pm 1$ and $T^a$ is a generator of the gauge algebra. It is clear that one is free to relabel $-s^aA'^a{}^\mu \to A'^a_\mu$, (for each gauge factor independently) without affecting $A^a_\mu$, which would turn the above P into CP, and reciprocally. This relabelling changes a given representation into its conjugate and flips the hypercharge, as it should for the action of charge conjugation. There are no physical consequences of either choice.\\
As for P, CP needs to send a RH fermion to a LH fermion and reciprocally. Therefore, a quark $Q$ must be sent to $Q^{\prime\, c}$,
\begin{equation}
Q \xrightarrow[]{CP\circ\mathbb{Z}_2} \gamma^0 Q^{\prime\, c}\,,
\end{equation} 
with the spacetime dependence left implicit. Imposing parity demands that $Q^{\prime}$ is in the $\overline{\mathbf{3}}$ of $SU(3)^{\prime}$, and in the $\mathbf{3}$ when CP is imposed. Indeed,
\begin{equation}
\begin{split}
\bar Q\gamma^\mu T^a Q G_\mu^a \xrightarrow[]{P\circ\mathbb{Z}_2}  &-\bar Q^{\prime}\gamma_\mu T^a{}^T Q^{\prime} G^\mu{}^a\\
&=\bar Q’\gamma_\mu \(-T^a{}^*\) Q’ G^\mu{}^a \ , \\
\bar Q\gamma^\mu T^a Q G_\mu^a \xrightarrow[]{CP\circ\mathbb{Z}_2}  &\ \bar Q’\gamma_\mu s^aT^a{}^T Q’ G^\mu{}^a\\
&=\bar Q’\gamma_\mu T^a Q’ G^\mu{}^a\,
\end{split}
\label{eq:PZ2CPZ2}
\end{equation}
(for $T^a=\lambda^a/2$, with $\lambda^a$ a Gell-Mann matrix), where one recognizes the generators in the $\overline{\mathbf{3}}$ as the opposite of the complex conjugates of those in the $\mathbf{3}$. More generally, the fermions charges for CP would be given by Tab.~\ref{tab:Spectrum} where all prime representations are conjugated. The two transformations in Eq.~\eqref{eq:PZ2CPZ2} can be exchanged upon renaming the gluons appropriately. The $\bar \theta$ angles are odd under CP as under P, and the condition on the Yukawa matrices are independent of the choice of P or CP, hence $\bar\theta = -\bar\theta^{\prime}$ and the discussion proceeds exactly as in the main text.

\section{Heavy Gluons}
\label{app:HeavyGluons}

The breaking of $SU(3)\times SU(3)^{\prime}\to SU(3)_{\rm QCD}$ results in 8 massless gluons and one heavy vector octet. The breaking being diagonal, implies that, up to a gauge transformation,
\beq
g'_3G_\mu^a=\pm g'_3G_\mu^{\prime a}
\eeq
once one projects onto the massless gluons, where the sign may depend on $a$. For instance, when one introduces the vev $\langle\Sigma\rangle=\frac{v_3}{\sqrt 6}\mathbf{1}$ of the bi-fundamental scalar $\Sigma$, its kinetic term contains
\beq
 \frac{v_3^2}{12}\left\{\begin{array}{l} (\mathbf{3},\mathbf{3}): 
 (g_3G^a_\mu+s^ag_3' G'^a_\mu)^2 \\
 (\mathbf{3},\overline{\mathbf{3}}): (g_3G^a_\mu- g_3'G'^a_\mu)^2\
 \end{array}\right. \ .
\label{eq:gluonMasses}
\eeq
Thus, the projection onto the massless gluons yields
\beq
g_3^2G\tilde G=g_3'^2G'\tilde G' \ ,
\eeq
which shows that the $\bar\theta$ term vanishes in the IR as discussed in the main text.
Choosing the realization where $\Sigma\in\(\mathbf{3},\mathbf{\bar 3}\)$ for definiteness, we can diagonalize the gluon matrix in Eq.~\eqref{eq:gluonMasses} to obtain the mass eigenstates $G_{SM}$ and $G_{H}$: 
\begin{align}
    G^{\mu\,a}_{SM} =& \frac{g_3^{\prime}G^{\mu\,a} + g_3G^{\prime\, \mu\,a}}{\sqrt{g_3^{\prime\, 2} + g_3^2}}\,,\\
    G^{\mu\,a}_{H} =& \frac{g_3G^{\mu\,a} - g_3^{\prime}G^{\prime\, \mu\,a}}{\sqrt{g_3^{\prime\, 2} + g_3^2}}\,.
\end{align}
Below $v_3$ one recovers the SM group of strong interactions, $SU(3)_{\rm QCD}$, plus a massive spin-one octet. The gauge-fermion interactions in the mass basis now reads:
\beq
\bead
\mathcal{L}\supset &- g_{\rm QCD}\bar{q}\gamma_{\mu}q T^a \(-\tan\theta G^{\mu\, a}_H + G^{\mu\, a}_{SM}\)\\
&- g_{\rm QCD}\bar{q}^{\prime}\gamma_{\mu}q^{\prime} T^a \(\tan^{-1}\theta G^{\mu\, a}_H + g_sG^{\mu\, a}_{SM}\)\,,
\eead
\label{eq:GSMGHcouplings}
\eeq
where we identified the QCD coupling constant $g_{\rm QCD}\equiv\frac{g_3g_3^{\prime}}{\sqrt{g_3^2+g_3^{\prime\, 2}}}$ and defined $\tan\theta\equiv g_3/g_3'$. Eq.~\eqref{eq:GSMGHcouplings} shows that a heavy gluon can decay to two light quarks.

\section{Physical Colored Scalars in $\Sigma$}
\label{app:Scalars}

Considering again the case of the bi-fundamental scalar field $\Sigma$, one finds that the physical scalars that it contains decompose into two real singlets of $SU(3)_{\rm QCD}$ and an octet, which have been thoroughly studied in Ref.~\cite{Bai:2018jsr}. In unitary gauge,
\beq
\Sigma=\frac{v_3+\sigma+i\tilde\sigma}{\sqrt 6}\mathbf{1}+\sigma^aT^a \ .
\eeq
From the most general potential of $\Sigma$,
\beq
\begin{split}
V(\Sigma) =& -m^2\Tr(\Sigma\Sigma^\dagger)+c\Tr^2(\Sigma\Sigma^\dagger)\\
&+\tilde c\Tr(\Sigma\Sigma^\dagger)^2+\(\tilde m\det(\Sigma)+h.c.\)
\ .
\end{split}
\label{eq:SigmaPotential}
\eeq
one finds
\beq
v_3=\frac{\sqrt 3 \left(\sqrt{8 (3 c+\tilde c)m^2+\tilde m^2}-\tilde m\right)}{2\sqrt 2 (3 c+\tilde c)}
\label{v3Value}
\eeq
and the interactions read, in the mass basis,
\beq
\bead
\mathcal{L} = &\frac{m_{0^+}^2}{2}\sigma^2+\frac{m_{0^-}^2}{2}\tilde\sigma^2+\frac{m_8^2}{2}\sigma^a{}^2\\
&+\(g_8+\lambda'_{0^+8}\sigma\)d^{abc}\sigma^a\sigma^b\sigma^c+g_{0^+}\sigma^3\\
&+\lambda_{0^+}\sigma^4+\lambda_{0^-}\tilde\sigma^4+g_{0^+0^-}\tilde\sigma^2\sigma+\lambda_{0^+0^-}\tilde\sigma^2\sigma^2\\
&+\lambda_8\(\sigma^a{}^2\)^2+\(g_{0^+8}\sigma+\lambda_{0^+8}\sigma^2+\lambda_{0^-8}\tilde\sigma^2\)\sigma^a{}^2 \ ,
\eead
\eeq
where $d^{abc}\equiv \frac{1}{2}\Tr\(T^a\[T^b,T^c\]\)$. The masses are 
\beq
\begin{split}
m_{0^+}^2 &= v_3^2\tilde{c}+ \frac{2v_3\tilde{m}}{\sqrt{6}} + 3\, c\, v_3^2 - m^2\,,\\
m_{0^-}^2 &= \frac{v_3^2\tilde{c}}{3} - \frac{2v_3\tilde{m}}{\sqrt{6}} + c\, v_3^2 - m^2\,,\\
m_8^2&=v_3^2\tilde{c} - \frac{v_3\tilde{m}}{\sqrt{6}} + c\, v_3^2 - m^2 \, .
\end{split}
\label{eq:colorMassesSigma}
\eeq
$\sigma^a,\tilde\sigma$ and the 8 longitudinal polarizations of $G_H$ are the 17 Nambu-Goldstone bosons (NGBs) of the $O(18)$ symmetry of the $\Sigma$ kinetic term, spontaneously broken to $O(17)$ by $\langle\Sigma\rangle$. It is partly gauged and partly explicitly broken, which can be seen from the tree level masses above once one inserts the value of $v_3$ as given in Eq.~\eqref{v3Value}. In the potential, only $\tilde m$ breaks the $U(1)$ symmetry of which $\tilde\sigma$ is the NGB, and its mass can consistently be seen to vanish in the $\tilde m\to 0$ limit. In this limit, the mass of $\sigma^a$ is found to be proportional to $\tilde c$, the only coupling left in the potential which breaks $O(18)$.\\
The trilinear couplings read
\beq
\begin{split}
g_{0^+} &= \frac{v_3\tilde{c}}{3}+\frac{\tilde{m}}{3\sqrt{6}}+v_3\,c\,,\\
g_8&=\frac{2\sqrt 2v_3\tilde{c}}{\sqrt 3}+\frac{2\tilde{m}}{3}\,,\\
g_{0^+0^-} &= \frac{v_3\tilde{c}}{3}-\frac{\tilde{m}}{\sqrt{6}}+v_3\,c\,,\\
g_{0^+8}&=v_3\tilde{c}-\frac{\tilde{m}}{2\sqrt 6}+v_3\,c\,,\\
\end{split}
\eeq
and the quartic couplings are
\beq
\begin{split}
\lambda_{0^+} &= \lambda_{0^-} = \frac{\tilde{c}}{12}+\frac{c}{4}\,,\\
\lambda_{8} &= \frac{\tilde{c}}{8}+\frac{c}{4}\,,\quad \lambda_{0^+8} = \frac{\tilde{c}}{2}+\frac{c}{2}\,,\\
\lambda_{0^+0^-} &= \frac{\tilde{c}}{6}+\frac{c}{2}\,,\quad \lambda^{\prime}_{0^+8} = \frac{4\tilde{c}}{\sqrt{6}} \, .
\end{split}
\eeq
From these couplings we see that $\sigma$ and $\sigma^a$ can decay to gluons via a loop of $\sigma^a$. (These new scalar bosons could also decay to other new bosons, depending on their masses.) Through the scalar potential, the P-odd $\tilde\sigma$ is stable, but it can decay to gluons and quarks via an intermediate $\sigma^a$ and $G_H$, to which it couples via the following interaction contained in the $\Sigma$ kinetic term,
\beq
\frac{\(1+\tan^2\theta\)g_s}{\sqrt 6\tan \theta}G_{H\mu}^a\tilde\sigma\overset{\leftrightarrow}{\partial^\mu}\sigma^a \ .
\eeq

\section{Dynamical Color Breaking}
\label{sec:composite}

In the main text, we argued that a small $\langle\Sigma\rangle/v'$ ratio makes the model as testable as possible. Nevertheless, from a theory perspective this may call for an explanation, as $\langle\Sigma\rangle$ suffers from a hierarchy problem. Here, we explore dynamical color breaking as an alternative. In addition to evading hierarchy problems, such scenarios can be characterized by a single additional parameter, while the $\Sigma$ potential has four. It can also be seen from the following discussion that the phenomenology is affected by the choice of color breaking mechanism. For instance, in the dynamical models, composite scalars are parametrically lighter than $v_3$ and do not have a tunable mass.\\
A tentative model which replaces the $\Sigma$ dynamics has the field and gauge content shown in Table~\ref{tab:composite1}.
\begin{table}[t]
    \centering
    \begin{tabular}{c|c|c|c}
        & $SU(N)$ & $SU(3)$ & $SU(3)'$ \\
        \hline
$\psi_L$& $\mathbf{N}$ & $\mathbf{3}$ & $\mathbf{1}$ \\
$\psi_R$& $\mathbf{N}$ & $\mathbf{1}$ & $\mathbf{3'}$ \\
    \end{tabular}
    \caption{Field and gauge content leading to a dynamical generation of $v_3$. Subscripts indicate chirality. $\mathbf{3'}$ refers to $\mathbf{3}$ or $\mathbf{\bar 3}$. Spectator fermions cancel the anomalies.}
    \label{tab:composite1}
\end{table}
The representation $\mathbf{3'}$ can be taken to be $\mathbf{3}$ or $\mathbf{\bar 3}$, mapping to $\Sigma$ in the $\(3,\bar 3\)$ or $\(3,3\)$, respectively. There are $SU(3)^{(\prime)3}$ anomalies, which must be cancelled by spectator fermions, to be discussed below. When the $SU(N)$ gauge group becomes strong and confines (which happens for $N\geq 2$, as there are three flavors), one expects a quark condensate to form and break the axial part of the flavor group \cite{Coleman:1980mx}. The full flavor group being gauged, all vacua are equivalent to the one such that
\beq
\langle\bar\psi_L\psi_R\rangle=v_3^3\mathbf{1}_{3\times 3} \ ,
\eeq
showing that the breaking pattern is the one announced, namely to the diagonal vectorlike $SU(3)_\text{QCD}$ subgroup. The eight NGBs of the broken $SU(3)_A$ flavor group are all eaten up by the heavy gluons, while the would-be NGB of the axial $U(1)_A$ receives a large mass from the mixed anomaly with $SU(N)$. A remark on the unbroken $U(1)_V$ is also in order: it has a mixed anomaly with both $SU(3)$ and $SU(3)'$, which shifts $\theta-\theta'$ but not $\theta+\theta'$. Therefore, parity in the UV is still the ingredient that solves the strong CP problem. How it acts depends on the precise structure of the model and will be discussed below.\\
The spectator fermions that cancel the gauge anomalies are colored, hence they must be made massive. We discuss different options below, which we classify according to whether they lead to a high or low $v_3$. 

\subsection{Models with high $v_3$}

The colored spectators can be given a mass through a 4-Fermi interaction with the fermions participating in the above condensate. Focusing on fundamentals of $SU(3)^{(\prime)}$, we consider $\psi'_{L/R,i=1...N}$, where $i$ is a flavor index, in the conjugate representation of $SU(3)^{(\prime)}$ with respect to $\psi_{L/R}$, respectively. We can then add to the lagrangian
\beq
\cL\supset \frac{c_{ij}}{M_\text{UV}^2}\overline{\psi_L^a}\psi_R^{a'}\overline {\psi_{R,ia'}^{\prime}}\psi_{L,ja}^{\prime} + h.c. \ ,
\eeq
where we made the contracted $SU(3)^{(\prime)}$ indices explicit. Due to the gauge structure of that model, parity must exchange $\psi_L\leftrightarrow \gamma^0\psi_R$ (hence we must choose $\mathbf{3'}=\mathbf{3}$), $\psi_L'\leftrightarrow \gamma^0\psi_R'$ and leave the gauge factor $SU(N)$ invariant. Therefore, $c=c^\dagger$ and $\bar\theta_\text{QCD}$ is not affected by the mass diagonalization of the spectators. In addition, the condensate, on which parity acts as hermitian conjugation, does not break parity. Finally, the theta angle $\theta_N$ of $SU(N)$ must vanish. Therefore, the would-be NGB of $U(1)_A$, which relaxes $\theta_N$ to zero and couples like $\theta+\theta'$, does not shift the latter as it does not acquire a vev.\\ 
By consistency, the cutoff $M_\text{UV}$ ought to be at least $v'$. Consequently, imposing that the new colored particles are heavier than a TeV leads to the following bound on $v_3$:
\beq
v_3\gtrsim 10^7\text{ GeV }\(\frac{M_\text{UV}}{v'}\)^{2\over 3}\(\frac{v'}{10^9\text{ GeV}}\)^{2\over 3} \ ,
\eeq
which is such that the heavy gluons cannot be seen at the LHC. Here, the lightest BSM particles which can be searched for are colored fermions (found both in the mirror and spectator sector).\\ 
Another option, free of any explicit cutoff, consists instead in charging the spectator fermions under an additional confining gauge theory. This leads us to the gauge and field content of Table~\ref{tab:composite2}.
\begin{table}[b]
    \centering
    \begin{tabular}{c|c|c|c|c}
        & $SU(N)$ &$SU(N)'$ & $SU(3)$ & $SU(3)'$ \\
        \hline
$\psi_L$& $\mathbf{N}$ & $\mathbf{1}$ & $\mathbf{3}$ & $\mathbf{1}$ \\
$\psi_R$& $\mathbf{N}$ & $\mathbf{1}$ & $\mathbf{1}$ & $\mathbf{3'}$ \\
$\psi_L'$& $\mathbf{1}$ & $\mathbf{N}$ & $\mathbf{\bar 3}$ & $\mathbf{1}$ \\
$\psi_R'$& $\mathbf{1}$ & $\mathbf{N}$ & $\mathbf{1}$ & $\mathbf{\bar 3'}$ \\
    \end{tabular}
    \caption{Field and gauge content extending that of Table~\ref{tab:composite1} and leading to a dynamical generation of $v_3$. Subscripts indicate chirality. $\mathbf{3'}$ refers to $\mathbf{3}$ or $\mathbf{\bar 3}$.}
    \label{tab:composite2}
\end{table}
The model is easily seen not to generate any gauge anomalies, as it is vectorlike with respect to each gauge factor independently. (It is chiral at the level of the whole gauge group, but all mixed anomalies are trivially zero.)\\
Two scenarios can be considered: the two $SU(N)$ groups confine at different or similar scales, depending on the charge assignments that determine the action of parity. When $\mathbf{3'}=\mathbf{3}$, parity must exchange $\psi_L^{(\prime)}\leftrightarrow \gamma^0 \psi_R^{(\prime)}$ and leave the $SU(N)$ factors unchanged, thereby making the gauge couplings and the confining scales $\Lambda$ and $\Lambda^{\prime}$ independent. When $\mathbf{3'}=\mathbf{\bar 3}$, parity must be chosen to exchange the two $SU(N)$, as well as $\psi_L\leftrightarrow \gamma^0 \psi_R',\psi_L'\leftrightarrow \gamma^0\psi_R$. This equates the two gauge couplings in the UV and relates the two confining scales, $\Lambda\approx \Lambda'$. (The equality may not hold exactly, due to higher-dimensional terms which lift the degeneracy between the $SU(3)^{(\prime)}$ gauge couplings below $v'$ at the classical level.) This second option is consistent with a low $v_3$ and will be discussed in the next section. In the first case, we can pick $\Lambda\geq\Lambda'$ without loss of generality, and the only possible condensates are
\beq
\label{vacuumDoubleSUN1}
\langle \overline{\psi_L}\psi_R\rangle={\cal V}_3^3\mathbf{1}_{3\times 3} \ , \ \langle \overline{\psi'_L}\psi_R'\rangle={\cal V}_3^{\prime 3}\mathbf{1}_{3\times 3} \ ,
\eeq
where $v_3 \approx{\cal V}_3\approx \Lambda/(4\pi)\geq {\cal V}'_3\approx \Lambda'/(4\pi)$. As said above, the first condensate can always be brought to that form through gauge transformations, while the second is selected by vacuum alignment \cite{Weinberg:1975gm,Susskind:1978ms,Peskin:1980gc,Preskill:1980mz} with the vectorlike QCD gauge group below $v_3$. The condensates do not break parity but break the (non-anomalous) $SU(3)_A^2$ flavor symmetry, whose $16$ NGBs decompose into the eight longitudinal components of the heavy gluons, and a colored octet of physical scalars $\pi^a$, of mass
\beq
\label{pionsDoubleSUN1}
m_{\pi^a}^2\sim \frac{\alpha_s}{4\pi}\Lambda'{}^2
\eeq
and phenomenology identical to that of the octet of scalars discussed above. Since there are collider bounds on such colored scalar octets (which, as seen above, can decay to gluons), $\Lambda'$ is bounded from below. When the $SU(N)^{(\prime)}$ coupling constants are not tuned to be very similar at the scale $v'$, we expect that $\Lambda'\ll v_3$. Therefore, as above, the heavy gluons are not accessible at the LHC, but the light NGBs can be looked for. As before, the unbroken anomalous $U(1)_V$ symmetries do not affect $\theta+\theta'$, while the two would-be NGBs, which relax to zero the already-vanishing $\theta_{N^{(\prime)}}$, do not get a vev and do not shift $\theta+\theta'$ either.

\subsection{Models with low $v_3$}

In the second case where $\mathbf{3'}=\mathbf{\bar 3}$ and $\Lambda\approx \Lambda'$, $v_3$ can be chosen to saturate the current experimental bounds, such that the heavy gluons can be in principle searched for. In that case, we note that the leading-order techniques of Refs.~\cite{Weinberg:1975gm,Susskind:1978ms,Peskin:1980gc,Preskill:1980mz} do not allow us to determine the precise vacuum and cannot rule out the possibility that $SU(3)_\text{QCD}$ is also partly broken by the vacuum. (In that case, parity would also be spontaneously broken. Due to the gauge invariance, the only relevant cases are unbroken $SU(2)\times U(1)$ and $U(1)^2$.) A definite answer is found upon analyzing the vacuum beyond leading order, as done e.g. in Refs.~\cite{Peskin:1980gc,Chadha:1981rw,Chadha:1981yt}. We find instabilities for all cases but the one where $SU(3)_\text{QCD}$ is conserved, for which the vacuum is again that of Eq.~\eqref{vacuumDoubleSUN1}. (Due to the identical confinement scales, it respects parity.) The physical spectrum is then again made of an octet of heavy gluons and an octet of NGBs. However, the masses that the latter receive from QCD are not given by Eq.~\eqref{pionsDoubleSUN1}, but arise at higher order, namely only at order $\cO(\alpha_s^2)$ (see e.g. \cite{Peskin:1980gc,Chadha:1981rw,Chadha:1981yt} for an explanation based on symmetries). More precisely, we find that the scalars receive a mass
\beq
\label{pionsDoubleSUN2}
m_{\pi^a}^2\sim \frac{\alpha_sm_G^2}{4\pi}\log\(\frac{\Lambda^2}{m_G^2}\) \ ,
\eeq
where the heavy gluon mass $m_G^2\sim \alpha_s v_3^2$. The NGBs are thus lighter (by roughly an order of magnitude) than the gauge bosons, and would constitute the first collider signals. Given current constraints on the NGBs, $v_3 \gtrsim\, \text{30 TeV}$ and the heavy gluons could be within the reach of future colliders, unlike in the previous section. Something non-trivial happens at the level of the would-be NGBs of the two $U(1)_A$: they now can relax the non-vanishing $\theta_{N^{(\prime)}}$, but those are opposite due to parity, hence the vevs of the two would-be NGBs are opposite. Since they couple identically to $G^{(\prime)}\tilde G^{(\prime)}$, they do not shift $\theta+\theta'$.
\\ 
In the case of a single $SU(N)$, we added $SU(N)$-singlet spectator fermions to cancel the color anomalies. Another possibility is to double the spectrum of strongly interacting fermions, so that the color anomalies are cancelled without spectators, as in the field content of 
Table~\ref{tab:composite3}.
\begin{table}[t]
    \centering
    \begin{tabular}{c|c|c|c}
        & $SU(N)$ & $SU(3)$ & $SU(3)'$ \\
        \hline
$\psi_L$& $\mathbf{N}$ & $\mathbf{3}$ & $\mathbf{1}$ \\
$\tilde\psi_L$& $\mathbf{N}$ & $\mathbf{\bar 3}$ & $\mathbf{1}$ \\
$\psi_R'$& $\mathbf{N}$ & $\mathbf{1}$ & $\mathbf{3}$ \\
$\tilde \psi_R'$& $\mathbf{N}$ & $\mathbf{1}$ & $\mathbf{\bar 3}$ \\
    \end{tabular}
    \caption{Field and gauge content extending that of Table~\ref{tab:composite1} and leading to a dynamical generation of $v_3$.}
    \label{tab:composite3}
\end{table}
Here, parity leaves the $SU(N)$ group unaffected and exchanges $\psi_L\leftrightarrow \gamma^0\psi_R'$, and similarly for $\tilde\psi^{(\prime)}$.\\
There are six flavors of fundamentals of $SU(N)$; thus, for $N\geq 2$, $SU(N)$ confines. The quark condensate forms and spontaneously breaks the $SU(6)_A$ flavor symmetry \cite{Coleman:1980mx}, part of which is weakly gauged. The largest possible unbroken piece of $SU(3) \times SU(3)'$ is $SU(3)_\text{QCD}$. As above, the leading-order techniques of Refs.\cite{Weinberg:1975gm,Susskind:1978ms,Peskin:1980gc,Preskill:1980mz} do not determine the actual vacuum, but a next-to-leading-order analysis sheds light on it: we found instabilities for selected situations where the gauge group breaks to $SU(2)\times U(1)$ or $U(1)^2$. Scanning over all possible vacuum alignments goes beyond the scope of this paper and is left for future work. Assuming $SU(3)_\text{QCD}$ does not break, two equivalent fermion condensates can form:
\beq
\langle \overline{\psi_L}\psi_R'\rangle=\langle \overline{\tilde\psi_L}\tilde\psi_R'\rangle=v_3^3\; \mathbf{1}_{3\times 3}
\eeq
with all other condensates vanishing, or
\beq
\langle \overline{\psi_L}\tilde\psi_R'\rangle=\langle \overline{\tilde\psi_L}\psi_R'\rangle=v_3^3\; \mathbf{1}_{3\times 3}
\eeq
with all other condensates vanishing. These realize the same breaking pattern as $\Sigma$ in the $\(3,\bar 3\)$ and $\(3,3\)$, respectively, and lead to $35$ NGBs $\Pi^A$ decomposing into $\mathbf{1}+\mathbf{3}+\mathbf{\bar 3}+\mathbf{6}+\mathbf{\bar 6}+\mathbf{8} +\mathbf{8}$ under the unbroken $SU(3)_\text{QCD}$. One octet is eaten by the gauge boson of $SU(3)_A$, and the would-be NGB of $U(1)_A$ gets a large mass $\sim v_3$, due to its mixed anomaly with $SU(N)$. As above, it does not get a vev and therefore it does not shift $\theta+\theta'$. The singlet NGB remains massless, protected by the broken generator $\text{diag}\(1,1,1,-1,-1,-1\)$ of $SU(6)_A$, which has no mixed anomalies with $SU(N)$ nor with $SU(3)^{(\prime)}$. As above, all the colored NGB only receive a mass from QCD at order $\cO(\alpha_s^2)$.

\bibliography{biblio}